\documentclass[10pt,conference]{IEEEtran}

\usepackage{graphicx}
\usepackage{arydshln}

\usepackage{url}
\usepackage{comment}

\usepackage{cite}

\usepackage{comment}
\begin{document}
\title{Cream Skimming the Underground: Identifying Relevant Information Points from Online Forums} 

%
%
\author{Felipe Moreno-Vera, Mateus Nogueira, Daniel S. Menasché, Miguel Bicudo,  Ashton Woiwood, Enrico Lovat, Anton Kocheturov, and Leandro Pfleger de Aguiar }

\newcommand{\myhl}[1]{}

\author{
\IEEEauthorblockN{Felipe Moreno-Vera\IEEEauthorrefmark{1}, Mateus Nogueira\IEEEauthorrefmark{1}, Cainã Figueiredo\IEEEauthorrefmark{1},  Daniel S. Menasché\IEEEauthorrefmark{1}, Miguel Bicudo\IEEEauthorrefmark{1}, \\ Ashton Woiwood\IEEEauthorrefmark{2},   Enrico Lovat\IEEEauthorrefmark{3}, Anton Kocheturov\IEEEauthorrefmark{3}, and Leandro Pfleger de Aguiar\IEEEauthorrefmark{4}}
\IEEEauthorblockA{
{\IEEEauthorrefmark{1}Federal University of Rio de Janeiro (UFRJ), }
{\IEEEauthorrefmark{3}Siemens Corporation, } \IEEEauthorrefmark{2}ESO, 
{\IEEEauthorrefmark{4}Amazon.com} 
}
}

\maketitle              

\newcommand\blfootnote[1]{%
  \begingroup
  \renewcommand\thefootnote{}\footnote{#1}%
  \addtocounter{footnote}{-1}%
  \endgroup
}

\thispagestyle{plain}
\pagestyle{plain}

\begin{abstract}
This paper proposes a machine learning-based approach for detecting the exploitation of vulnerabilities in the wild by monitoring underground hacking forums. The increasing volume of posts discussing exploitation in the wild calls for an automatic approach to process threads and posts that will eventually trigger alarms depending on their content. To illustrate the proposed system, we  use the CrimeBB dataset, which contains data scraped from multiple underground forums, and develop a supervised machine learning model that can filter threads citing CVEs and label them as Proof-of-Concept, Weaponization, or Exploitation.  Leveraging random forests, we indicate that accuracy, precision and recall above 0.99 are attainable for the classification task.  Additionally, we provide insights into the difference in nature between weaponization and exploitation, e.g., interpreting the output of  a decision tree,  and analyze the profits and other aspects related to the hacking communities.   Overall, our work sheds insight into   the exploitation  of vulnerabilities in the wild and can be used to provide additional ground truth to   models such as EPSS and Expected Exploitability.
\end{abstract}

\section{Introduction}

The exploitation of vulnerabilities in the wild poses significant threats to  the Internet ecosystem,  being a concern to end users, companies, and, more generally, to the stability of the Internet itself.  In essence, exploitation refers to the   use of a weaponized exploit to attack a target. In this stage, the attacker uses the weapon to take advantage of a vulnerability and gain unauthorized access to a system or steal sensitive information.    Therefore, early detection of weaponization and tentative exploitation is key for defending against attacks.\blfootnote{This paper appears at  IEEE International Conference on Cyber Security and Resilience (IEEE CSR), 2023. The first two authors contributed equally to the work. Corresponding authors: \{felipe.moreno.vera, msznogueira\}@gmail.com, sadoc@dcc.ufrj.br} 




While  public databases on weapons, such as ExploitDB, are  continuously updated with information about how to exploit vulnerabilities,  
underground hacking forums still contain privileged and more up-to-date  information about the availability and development of exploits and, more importantly, about the tentative use of those exploits in the wild~\cite{basheer2021threats,campobasso2022threat,pastrana2018crimebb}.  In particular, certain forums contain 
information about the prices of exploits, and   instructions on how to make attacks Fully UnDetectable (FUD). 
In this context, figuring out  what users are discussing in those forums is instrumental to detecting and neutralizing  the exploitation of vulnerabilities in the wild. 
Monitoring the discussion in these forums also allows for   tracking   exploit     prices,  their usage, demand, and  main targets.   

%

We use the CrimeBB dataset, made available by Cambridge Cybercrime Centre, which contains data scraped from multiple underground forums~\cite{pastrana2018crimebb}. 
We focus on activity related to hacking, noting that the increasing volume of posts discussing exploitation in the wild calls for an automatic approach to process threads   that will    trigger alarms depending on their content. To that aim,  we developed a supervised machine learning model, which filters threads citing a Common Vulnerabilities and Exposures  (CVE) identifier  and labels them as Proof-of-Concept (PoC), Weaponization, or Exploitation.  Then, we indicate rules that can be automatically derived from data, providing insights into the difference  between weaponization and exploitation.   

\textbf{Prior art.  } 
Weaponization and exploitation are two of the key stages involved in the development of a cyberattack.  Most of the literature has focused on weaponization~\cite{hanks2022recognizing,allodi2017economic}, i.e., the process of building exploits for vulnerabilities. Much less attention has been given to exploitation in the wild, i.e.,  the actual use of a weaponized exploit to attack a target, or to   gain unauthorized access into a system or steal sensitive information~\cite{basheer2021threats}.  In part, this occurs because the study of exploitation in the wild involves sensitive data and stringent non-disclosure agreements. 

Exploit Prediction Scoring System (EPSS)~\cite{jacobs2021exploit}  and Expected Exploitability~\cite{suciu2022expected}  are two examples of systems that aim at determining exploitability in the wild.  Whereas EPSS uses private sources to derive its parameters, Expected Exploitability uses public artifacts. However, to the best of our knowledge, there is no prior work using CrimeBB for the purpose of understanding exploitation in the wild.  Our work serves to close this gap and can be used to provide additional ground truth to previous models such as EPSS and Expected Exploitability. 

\textbf{Contributions.} In summary, our key contributions are  twofold. First, we provide an  analysis of exploitation of vulnerabilities in the wild, using the CrimeBB dataset.  We   conduct a longitudinal  analysis of profits and other aspects related to the hacking communities, e.g., indicating the prices associated with exploits and the   distribution of delays between discussions about vulnerabilities on those forums and the release of information at   the National Vulnerability Database (NVD).  Second, we present     a classifier for assessing eminent threats based on  underground forums.  

\textbf{Paper structure. } The remainder of this paper is structured as follows.  
In Section~\ref{sec:related} we discuss related work and  Section~\ref{sec:dataset} presents our dataset, with some general statistics. 
Section~\ref{sec:results} reports our empirical findings, in Section~\ref{sec:classifier} we discuss our thread classifier, and Section~\ref{sec:conclusion} concludes.   


\section{Related Work and Background} \label{sec:related}

In what follows,  we discuss related work  and background pertaining to the main themes of our work.

\subsection{NLP and threat intelligence (TI)}

The use of Natural Language Processing (NLP) for the analysis of hacker forums has been considered  in~\cite{rahman2021attackers,pastrana2019measuring,pastrana2018crimebb}.  In this work, we complement such body of literature by focusing on discussions about  software vulnerabilities within CrimeBB forums, which have been previously  considered for the analysis of eWhoring~\cite{pastrana2019measuring} and other  cybercrimes~\cite{pastrana2018crimebb}. 



Threat Miner~\cite{deguara2022threat} is a system to identify threats based on hacker forums. 
The authors of  Threat Miner classify notifications or reports   as ``good'' if they represent a cyber threat that can   be linked to a known CVE.  
In this study, we focus specifically on analyzing threads within CrimeBB forums that can be linked to known CVEs. 
By leveraging CVEs, we  relate data from CrimeBB forums against other sources such as  the Common Vulnerability Scoring Systems  (CVSS)  and EPSS   to gain new insights into the lifecycle of vulnerabilities.





\subsection{Blackhat forums} 
 Blackhat forums comprise unstructured posts.  All posts  include their content, author, and subject.  
Leveraging a public dataset collected by the CrimeBB project, we analyze  posts from multiple blackhat forums. 
In the posts, we find references to vulnerabilities, IP addresses, and products that are being exploited in the wild. 
Blackhat forums provide a way for researchers as well as badly-intentioned users to trade knowledge about hacking. These forums supply information ranging from beginner hacking skills to functional hacking tools that anyone can easily get access, sometimes  for free.    The so-called \emph{script-kiddies}, i.e., home users with limited computing skills, for instance, can leverage  those tools to initiate cyberattacks.   One of our goals is to distinguish between research activity that poses a potential threat against exploitation in the wild, wherein criminals chat about threats.   

\begin{figure}[t!]
    \centering
    \includegraphics[width=\columnwidth]{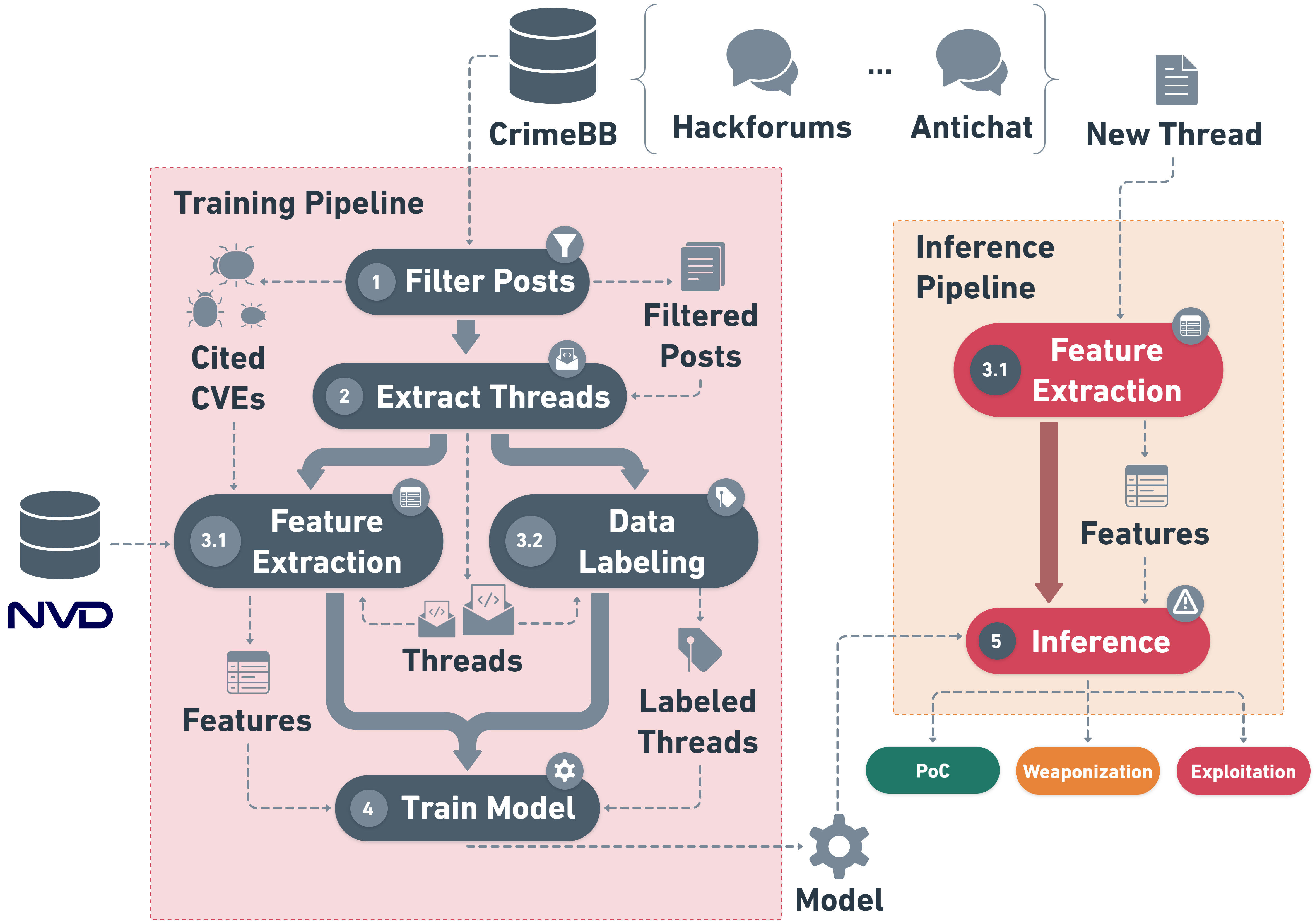}
    \caption{Proposed framework composed of two main steps: (1) threads and posts pre-processing including feature extraction and labeling; (2) three-class classifier to classify threads based on their content.}
    \label{fig:methodology}
\end{figure}

\section{Dataset} \label{sec:dataset}
Cambridge Cybercrime Centre makes available  sixteen  underground forums through CrimeBB. 
In these CrimeBB underground forums, we have 54,460,134 posts under 5,270,587 discussion threads. Those posts were filtered to extract information about software vulnerabilities.
In this work, we will focus on the largest forum, Hackforums, which has many boards for exchanging hacking knowledge, encompassing topics ranging from hacking to games. 
 In what follows, we provide further details on the dataset   used in this paper (Table~\ref{tab:table_posts_citing_cves}). \footnote{All the material to reproduce the results presented in this paper is available at \url{https://tinyurl.com/crimebbpaper}}

\subsection{Producing  the dataset}

To produce the dataset, we consider the following steps listed in Figure~\ref{fig:methodology}. First, we filter all posts citing at least one CVE (details in Section~\ref{sec:genstats}).  Each of those posts is contained in a thread.  
Then, we group all the posts in each of these threads, along with the thread title.  Finally, for each thread, we proceed with  feature extraction. The features correspond to the presence  of  words in   threads, e.g., through Bag-of-Words (BoW), Term Frequency-Inverse Document Frequency (TF-IDF), or doc2vec  
(Section~\ref{sec:features}). 
Then, our classifier takes all  features extracted from each thread as input and classifies the  thread into one of the considered target classes.   By classifying threads as opposed to individual posts, the proposed approach provides  a greater amount of contextual information to the classifier, which enhances classification accuracy.

 \setlength{\tabcolsep}{1.8pt}

\begin{table}[t]
    \centering
        \caption{Number of posts (threads) citing CVEs in the top 10 Hackforums  boards,  ranked by number of tagged posts} 
           \scalebox{0.95}{ 
    \begin{tabular}{|l|rrrrrr|rr|}
        \hline 
        Board & \multicolumn{8}{c|}{ Number of posts  (threads) citing CVEs } \\
        \cline{2-9}
         & \multicolumn{6}{c|}{Posts tagged  as   } & \multicolumn{2}{c|}{ } \\
        \cline{2-7} 
        & \multicolumn{2}{c}{PoC} & \multicolumn{2}{c}{Weapon} & \multicolumn{2}{c|}{Exploit} &  \multicolumn{2}{c|}{All  posts} \\
        \hline \hline
        Pentesting and Forensics & 271 & (55)  & 210 & (57) & 11 & (3) & 557 & (166) \\
        Premium Tools and Programs & 198 &(1)  & 28 & (3) & 142 & (4) & 433 &(20)\\
        Website and Forum Hacking & 93 &(34) & 139 & (43) & 16 & (12)   & 333& (132)\\
        Hacking Tools and Programs & 10 &(7) & 57 & (28) & 174 & (7) & 260& (59)\\
        Premium Sellers Section & -- & -- & 81 & (28) & 89 & (26) & 210 &(66)\\
        Beginner Hacking & 86 &(43) & 58 & (47)  & 6 & (6) &  219& (143)\\
        Botnets, IRC, and Zombies & 24 & (4)  & 85 & (34) & 22 & (5)  & 160 &(62)\\
        Hacking Tutorials & 58 &(21) & 8 & (4) & 3 & (3) &  74 &(33)\\
        Secondary Sellers Market & 8 &(4) & 33 &  (21) & -- & -- &  91& (40)\\
        News and Happenings & 9 &(9)  & 11 & (5) & 1 & (1) &   75 &(54)\\
        \hline \hline 
        Total,   all boards & 757 & (244) & 710 & (397) & 464 &(102)  & 3,037 &(1,162) \\
        \hline
    \end{tabular} }
    \label{tab:table_posts_citing_cves}
\end{table}

\subsection{Target classes and manual labeling}

Our search for vulnerabilities involved using a case-insensitive regular expression \texttt{cve-[0-9]\{4\}-[0-9]\{4,\}} (slightly more specific than \texttt{cve(-id)?(?i)} used  in~\cite{allodi2017economic}) to search for posts referring to vulnerabilities by their CVE identifiers. Across all CrimeBB forums, we found 4,116 posts citing 1,498 unique CVEs, under 1,700 discussion threads. 
This aligns with previous research on marketed exploits, which   considered similar quantities of posts~\cite{ablon2014markets,kotov2013anatomy,allodi2014comparing}. 
We discard the second most relevant forum, Antichat, due to the posts being primarily in Russian. We highlight that Antichat provides citations to around 90 additional unique CVEs. For Hackforums, we found 3,037 posts explicitly referring to 1,068 unique CVEs, under 1,162 discussion threads  between December 2007 and October  2019 (to be contrasted against 194 discussion threads and around 3,000 posts considered in~\cite{allodi2017economic}). 
From these 1,162 threads, a total of 1,067 were manually labeled by experts. \myhl{A subset of XXX posts was independently labeled by two experts, achieving    an inter-rater  agreement score of XXX. } The experts used the following code book to manually label the  threads:\footnote{Accounting for slang and abbreviations that are typical in those communities is left as a subject for future work.}

\begin{figure}[t!]
    \centering  \includegraphics[width=0.9\columnwidth]{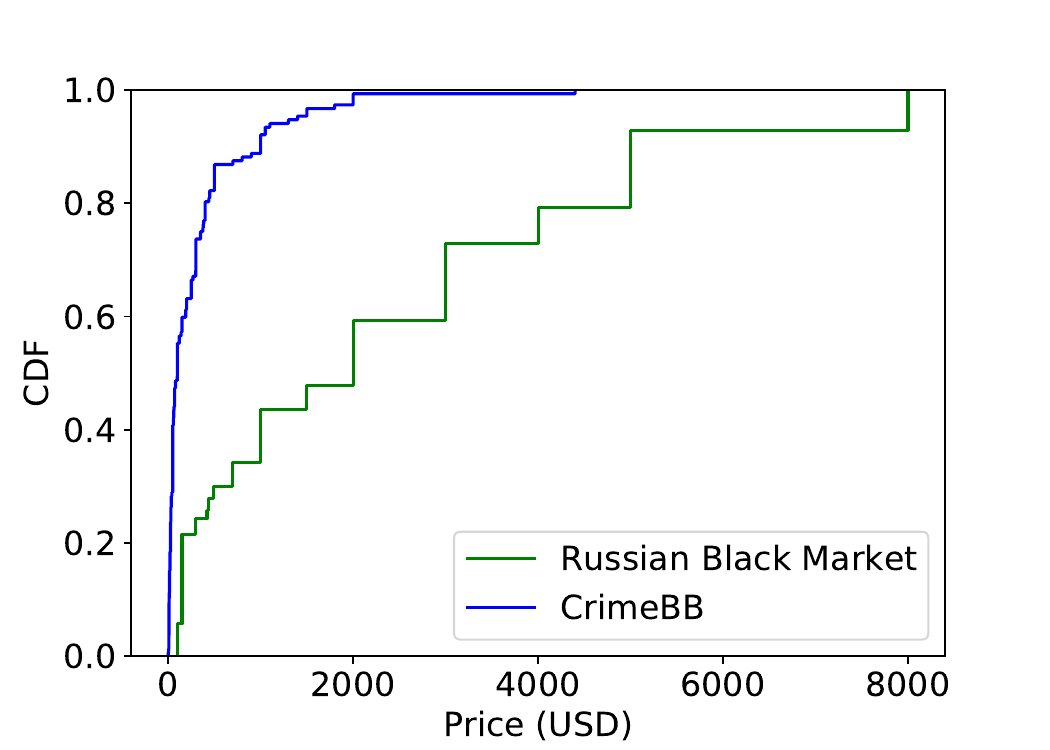}
    \caption{CDF of hacking tools prices:   prices at CrimeBB are relatively low compared against the Russian market -- some prices correspond to subscriptions, and others to  repackaging and FUD. Price statistics: CrimeBB (Min: 1, Median: 100, Max: 4400), Russian market (Min: 100, Median: 2000, Max: 8000)}
    \label{fig:cdf_prices}
\end{figure}

\begin{itemize}
    \item \textbf{PoC: } (1) contain keywords such as PoC, tutorial, guide (given the appropriate context of producing tools in a lab or controlled environment); (2) provide a tutorial description about how to build a PoC or (3) discuss vulnerabilities without signs of using exploits in the wild.   
    

    \item \textbf{Weaponization: } (1) contain keywords such as vulnerability and exploit (given the appropriate context of weaponization); (2) discuss the availability of fully functional or highly mature exploits, providing references or source code.
    

    \item \textbf{Exploitation: } (1) mention a well-known hacker group;  (2) contain references to cryptocurrencies and keywords such as bitcoin, exploitation, and attack (given the appropriate context of attacks in the wild); (3) discuss approaches to make exploits fully undetectable;  or  (4) involve markets  of exploits.
\end{itemize}
In addition to the above categories, the experts also labeled a few threads as \textbf{Scam}, when it was identified the selling of  an exploit  that was a posteriori  recognized as non-functional.  A total of 244, 397, 102 and 10 threads were labeled as PoC, Weaponization, Exploitation and Scam, respectively (see last line of Table~\ref{tab:table_posts_citing_cves}). Note that the remainder 314 threads  did not fit into any of the above categories. 
 Scams and the latter threads  were not considered in this study.

\begin{figure}[t!]
    \centering
   \includegraphics[width=0.9\columnwidth]{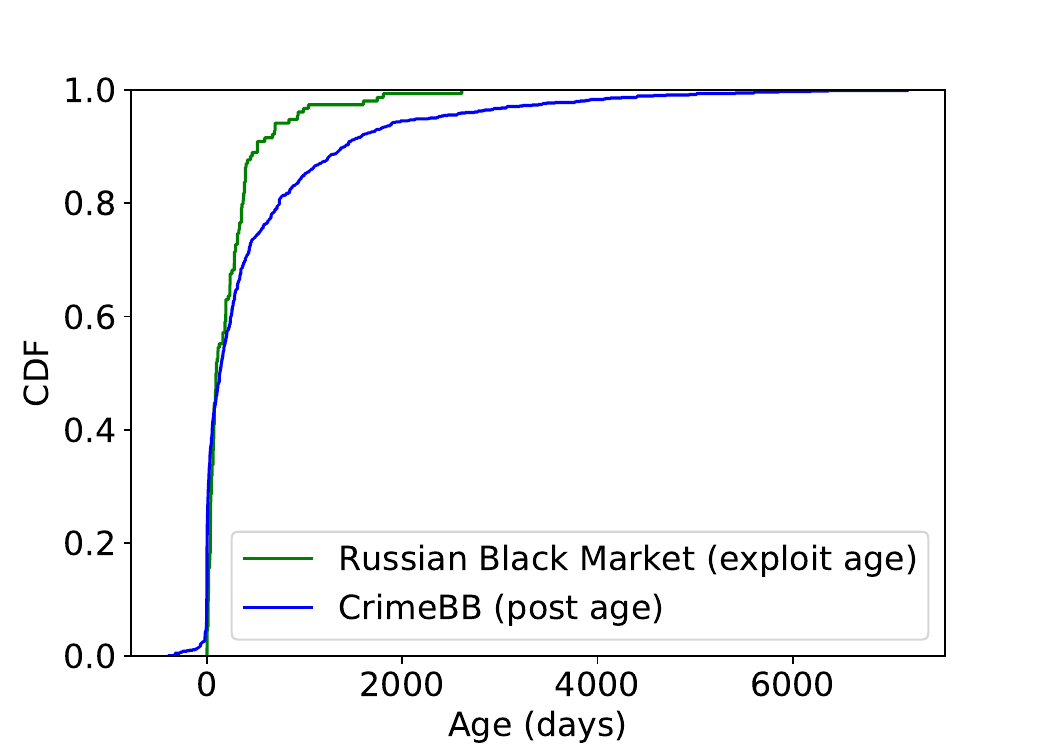}
    \caption{CDF of the difference in  days between CrimeBB citation   and NVD publish date. Negative values correspond to citations  to CVEs that occurred before NVD published the corresponding vulnerability. Age statistics: CrimeBB (Min: -396, Median: 132, Max: 7181), Russian market (Min: 1, Median: 95,5, Max: 2610) }
    \label{fig:cdf_days_crimebb_nvd}
\end{figure}

\subsection{Blackhat forums and markets statistics}

\subsubsection{General statistics} \label{sec:genstats}


In Table \ref{tab:table_posts_citing_cves}, we show the boards that contain most of the posts citing CVEs. In the top two boards, we find users selling and buying exploits, which indicates that discussions about vulnerabilities are generally about exploits already available on the market.  
%
Furthermore, Table~\ref{tab:table_posts_citing_cves} also shows the distribution of posts across different classes over the different boards at Hackforums. Note that in the board of pentesting, for instance, we find significant activity related to weaponization and exploitation. In contrast, few posts explicitly cite CVE identifiers in the board of hacking tutorials.

\subsubsection{Features} \label{sec:features}  The blackhat forums are unstructured.  The intrinsic features contained in the posts are the words that can be encoded using different strategies such as BoW, TF-IDF, and doc2vec.  In this work, we compare the three encoding techniques, noting that BoW is more interpretable, whereas TF-IDF and doc2vec yield higher accuracy (see Section~\ref{sec:results}).~\footnote{In future work, we consider leveraging additional features,  such as CVSS and EPSS scores of vulnerabilities and prices of   hacking tools.} 

\subsubsection{NVD data} We use data from  NVD to determine   properties of the considered vulnerabilities, such as severity level (CVSS). In particular,  NVD provides a brief description of each vulnerability, together with  its publish date, products affected, and external resources.

\section{Empirical Findings} \label{sec:results}

In this section, we report empirical findings from CrimeBB forums, including exploit prices, delays, and  risks.

\subsection{Prices} \label{sec:prices}

We discuss the prices  of artifacts cited by users from CrimeBB forums. Figure~\ref{fig:cdf_prices} shows the CDF of prices in dollars. For comparative purposes, we also plot the CDF of prices of exploits reported at the Russian market studied in~\cite{allodi2017economic}. Whereas in CrimeBB forums the minimum, median and maximum values were 1 USD, 100 USD, and 4,400 USD, in the Russian market the corresponding values were 100, 2,000, and 8,000~USD, respectively. Besides, we note that more than 80\% of the references to hacking tools correspond to prices  less than 1,000~USD. The larger prices observed in the Russian market when compared to CrimeBB forums can be explained by the fact that the Russian market requires explicit admission by market administrators.  Indeed, admission to the market is conditioned on  the user being active in related communities~\cite{allodi2017economic}. For this reason, in the Russian market users tend to discuss more mature, hence more expensive, artifacts. 

In the CrimeBB forums, in contrast, we observed that users tend to propose the repackaging of already existing exploits, e.g., under new FUD versions~\cite{valeros2020growth}.  Alternatively, some of the prices refer to subscriptions to websites that tend to be naturally cheaper than exploits. Despite the differences between prices, we also observe some similarities. In both platforms, the maximum prices did not surpass USD 8,000, the majority of prices are below USD 2,000, and roughly 20\% of the prices are close to USD 100.  Together, those numbers indicate that the activity in those forums can be monetarily rewarding, with rewards aligned with most bug bounty programs that offer up to  USD 3,000 for a critical bug.  Nonetheless, those numbers are still far from the million-dollar bug bounties that were recently reported in the literature.\footnote{\url{https://portswigger.net/daily-swig/million-dollar-bug-bounties-the-rise-of-record-breaking-payouts}}

%
%


\subsection{Delays} \label{sec:delays}

Knowing how long it takes for information about vulnerabilities to appear on online forums   is key, e.g., to assess risks associated with vulnerabilities and for patch management purposes~\cite{figueiredo2023statistical,miranda2021flow}. In this section, we  explore the delay between the publication of vulnerabilities at NVD and posts appearing at the forums. 
%
%
%
For CrimeBB forums, we compute the post age as the difference between the day of the post and the day on which the corresponding vulnerability was published at NVD, \texttt{PostAge = PostPubDate - CVEPubDate}.  Similarly, the exploit age reported by~\cite{allodi2017economic} is the difference between the day on which an exploit was published at the Russian market and the day on which the corresponding vulnerability was published at NVD, \texttt{ExplAge = ExplPubDate - CVEPubDate}. Figure~\ref{fig:cdf_days_crimebb_nvd} shows the CDF of \texttt{PostAge} and \texttt{ExplAge}, for CrimeBB forums and the Russian market, respectively.

Note that more than 50\% of exploits discussed in CrimeBB are about vulnerabilities that were disclosed over the previous  69 days before being cited at CrimeBB. Considering that 50\% of Industrial Control Systems (ICS) are not patched 60 days after vulnerability disclosure\cite{wang2017characterizing}, the use of blackhat forums is imperative to estimate risks associated with  vulnerabilities. Among the similarities between CrimeBB forums and the Russian black market, we observe that roughly 60\% of the  activity occurs very close to CVE publish date.  Discussion tends to phase out for virtually all vulnerabilities 6 years after they are released.  


We observe that in the Russian market, we have only positive age values, whereas under CrimeBB we have a small fraction of negative values. This is explained by the different nature of the two forums, as discussed in the previous section: whereas CrimeBB also counts with messages querying about vulnerabilities and discussing strategies to produce proof-of-concept weapons, the Russian market contains mostly discussion of mature exploits to be sold at higher values, and typically being released only after the CVE has already been published at NVD.  
 Fully functional or high-maturity exploits are rarely produced before the vulnerability   publish  date, i.e., \texttt{ExplPubDate} is larger than \texttt{CVEPubDate}.   Discussions about vulnerabilities, however, can initiate before they are released, as CVE identifiers are announced to the public before they are published at NVD. 


\subsection{Risks}
\label{sec:risks}
 
\begin{figure}[t!]
    \centering
\begin{tabular}{c}
 \includegraphics[width=\columnwidth]{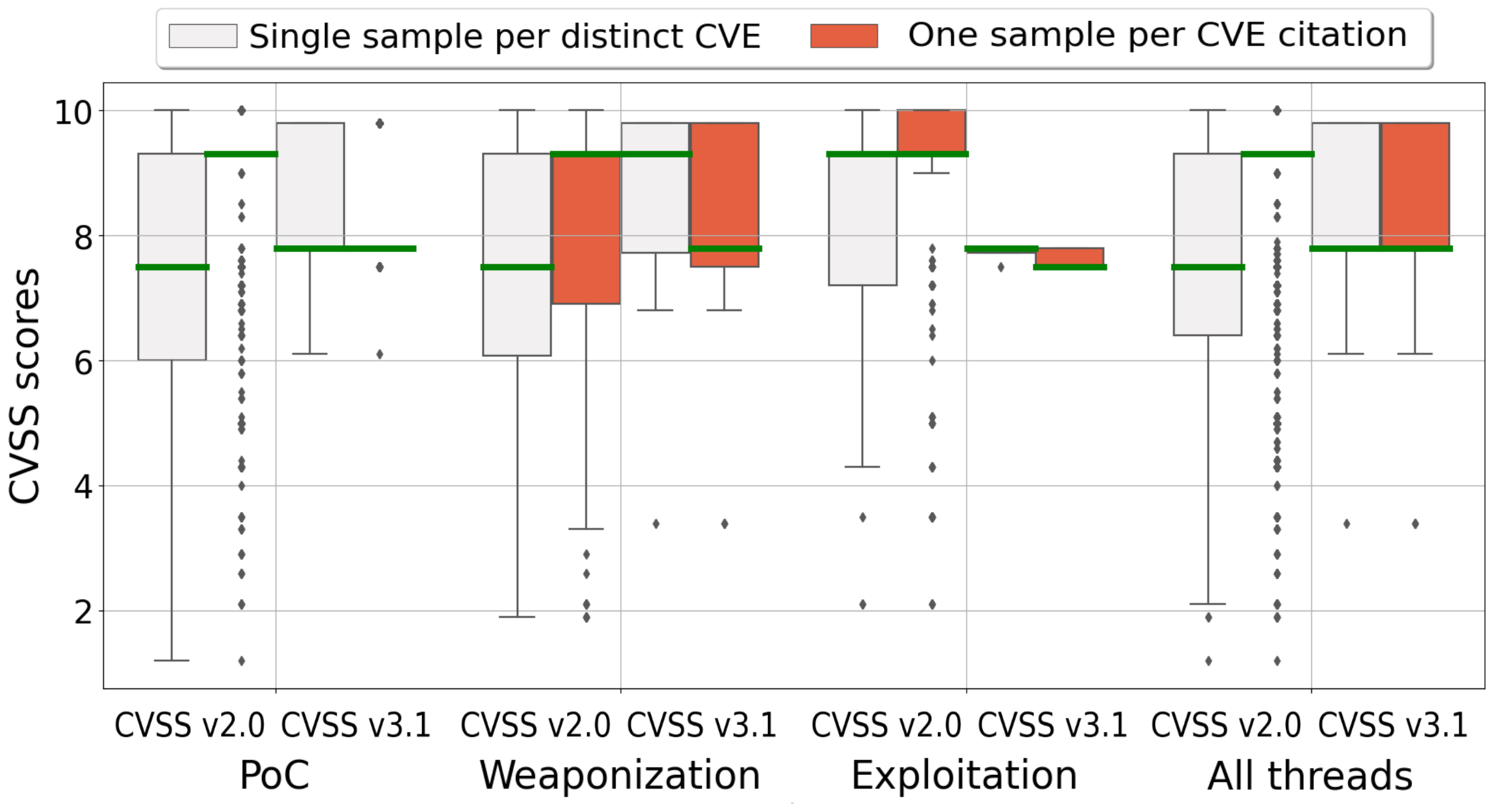}  \\
 (a) CVSS   \\
\includegraphics[width=\columnwidth]{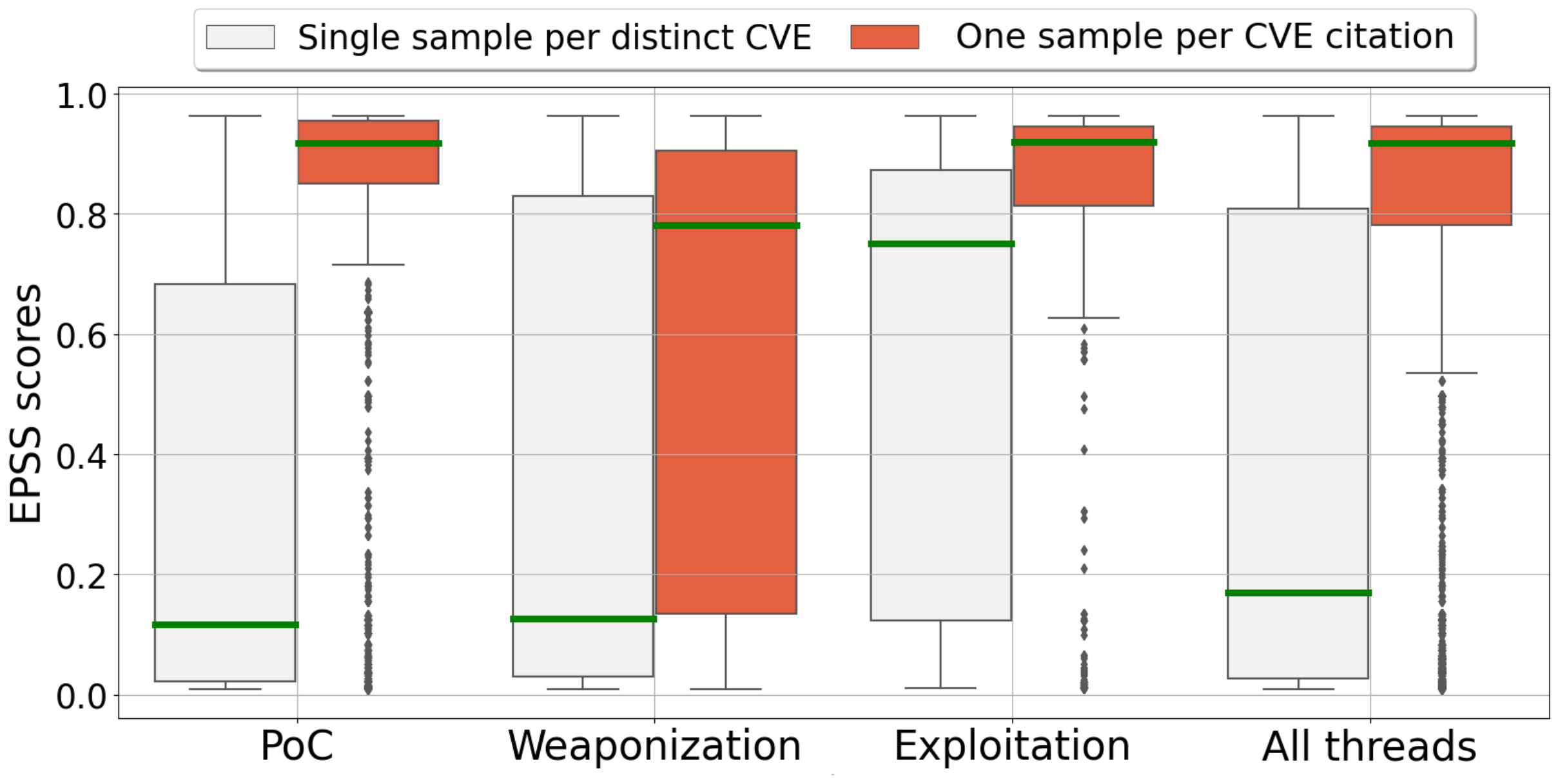} \\
(b) EPSS
\end{tabular}    
    \caption{Distribution of CVSS and EPSS scores across different classes.  Note that 91\% of posts refer to CVEs whose CVSS score is higher than the mean CVSS across all       NVD CVEs (not shown in the figure). } \vspace{-0.1in}
    \label{fig:cvss_epss_scores}
\end{figure}


Across all vulnerabilities, Figure~\ref{fig:cvss_epss_scores} shows the distribution of CVSS and EPSS scores conditioned on the thread class (PoC, Weaponization and Exploitation) and across all threads. Gray boxplots  consider one sample per CVE citation (counting repeated citations to a CVE within posts or threads as distinct citations), and red boxplots  consider  a single sample per distinct CVE (without counting repetitions). Figure~\ref{fig:cvss_epss_scores}(a) displays the distribution of  CVSS scores for CVSS versions 2.0 and 3.1  (the latter   is   available for a subset of CVEs). The overall median CVSS value exceeds 7.0.  Accounting for CVSS 2.0,  exploitation usually corresponds to higher CVSS values when compared against PoC and weaponization.  
 Additionally, still  accounting for CVSS 2.0,  CVEs with higher CVSS values are the most frequently cited. This is evidenced by the fact that the medians of the red boxplots are all above 9.0, indicating that the risk scores are magnified by repeated citations. A similar trend is observed in Figure~\ref{fig:cvss_epss_scores}(b)  for EPSS. EPSS was released in 2021, and we used its latest version, as of  February 28, 2023,  to represent the probability   of exploitation in the wild in the next 30 days.  Despite the gap between the post citations and the release of EPSS scores, we observe that EPSS scores are able to capture   high risks for the vulnerabilities   in our dataset.

\begin{figure}[t!]
    \centering
\begin{tabular}{c|c}
 \includegraphics[width=0.5\columnwidth]{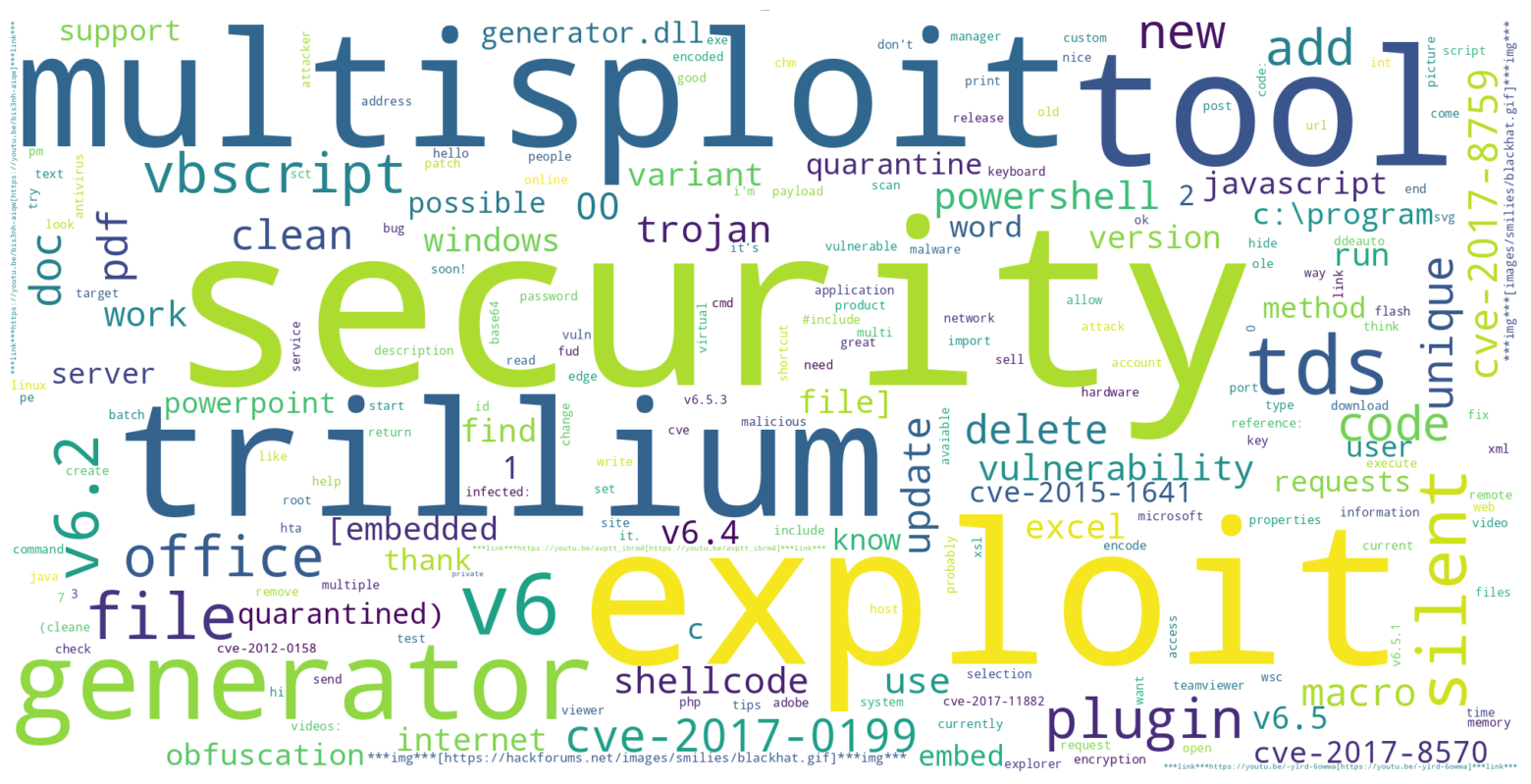}    & \includegraphics[width=0.5\columnwidth]{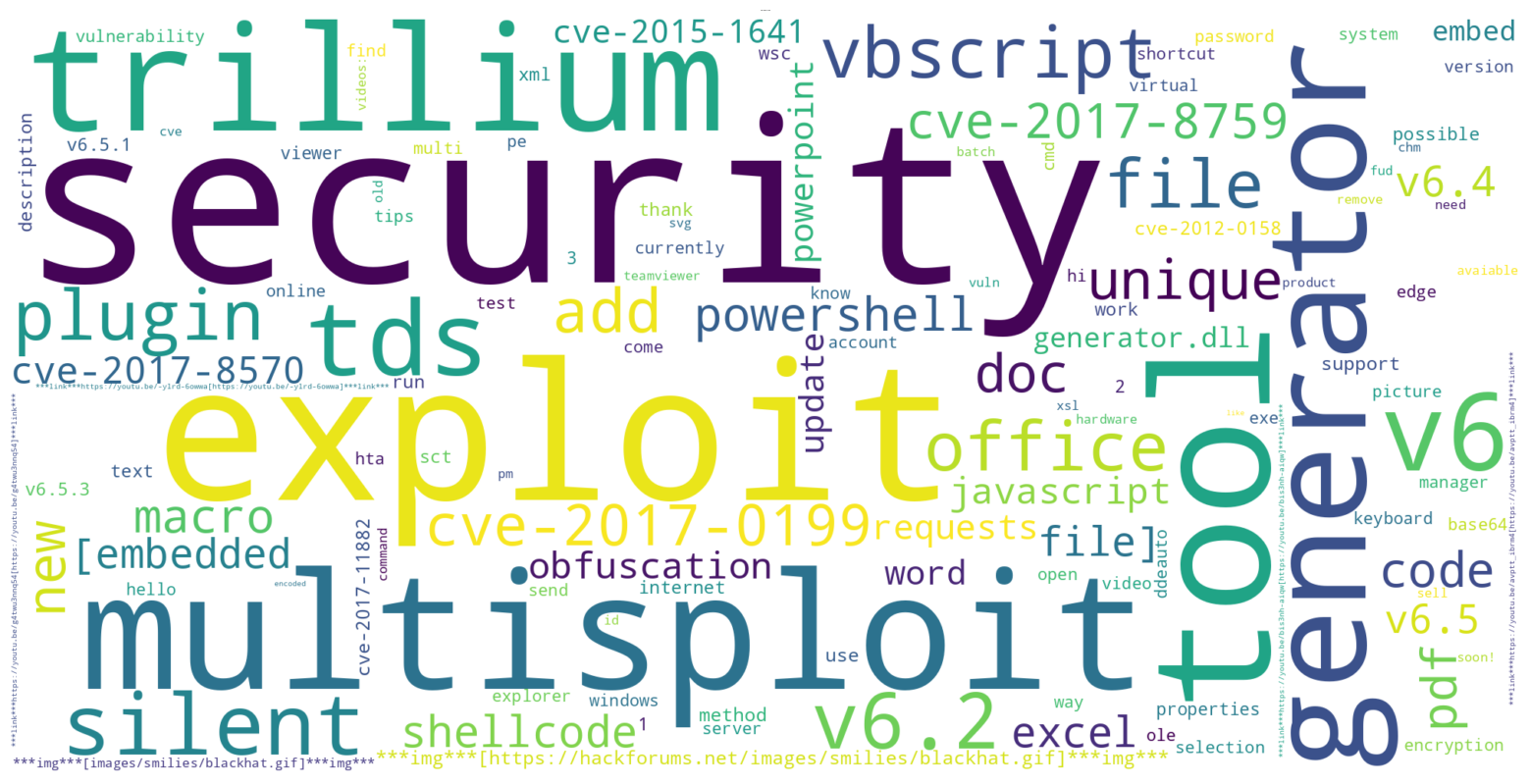}  \\
 (a) All  & (b) PoC \\
\includegraphics[width=0.5\columnwidth]{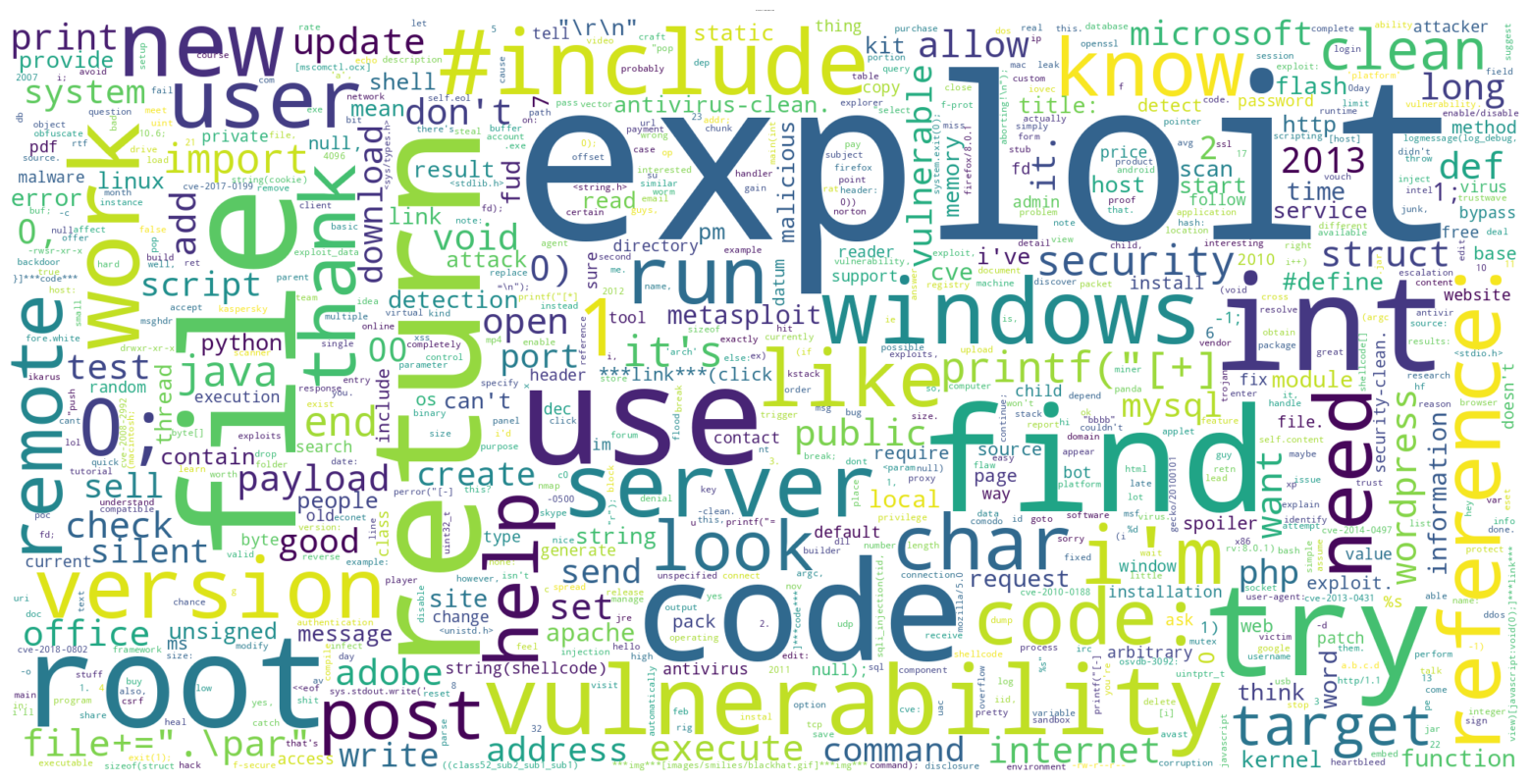}    &  \includegraphics[width=0.5\columnwidth]{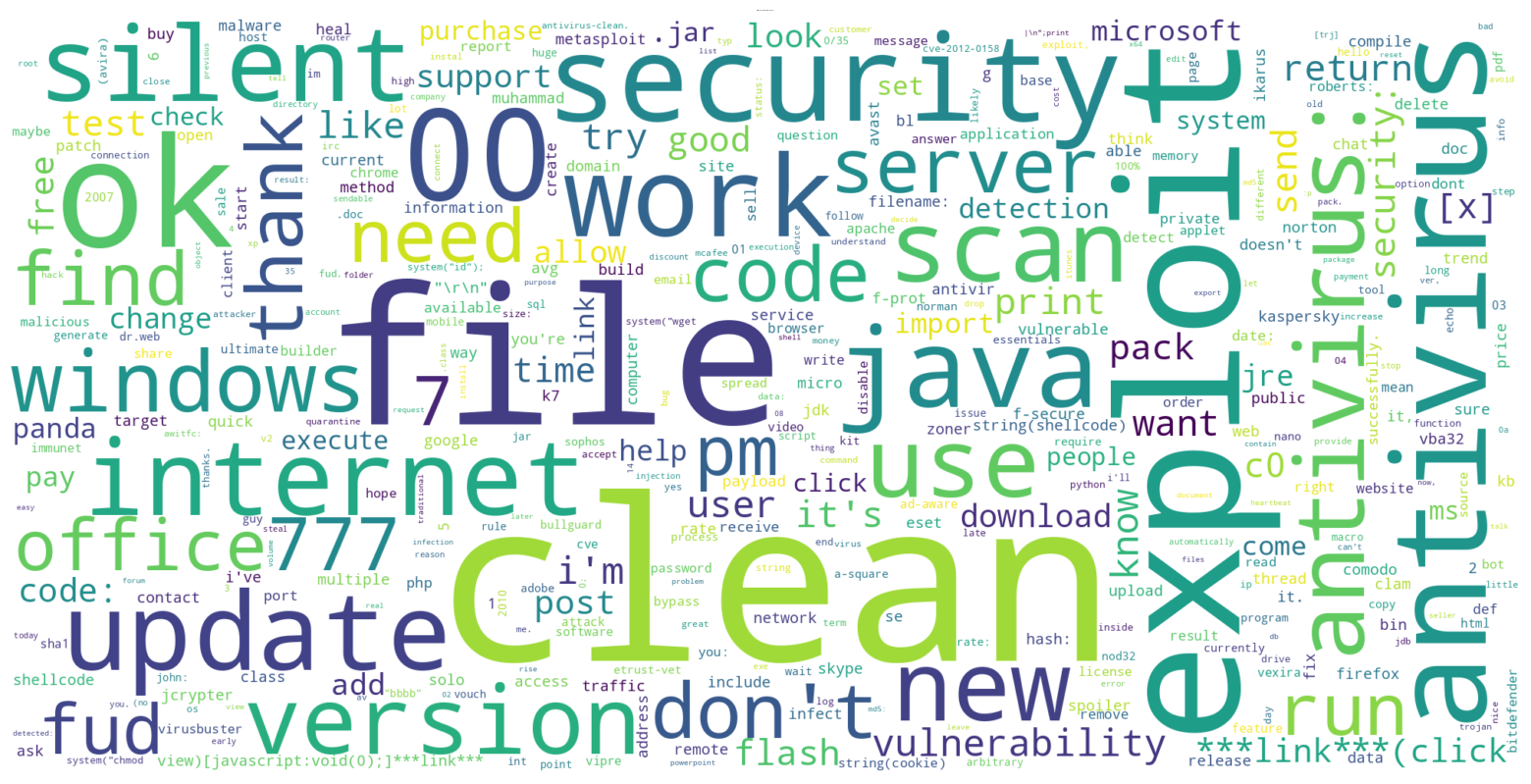} \\
(c) Weaponization & (d) Exploitation 
\end{tabular}    
    \caption{Word clouds showing the most frequent keywords appearing across posts: (a) all posts; (b) PoC; (c) weaponization and (d) exploitation. 
    }
    \label{fig:wordclouds} \vspace{-0.1in}
\end{figure}
 

\subsection{Natural Language Processing (NLP) and  word clouds}

Figure~\ref{fig:wordclouds} shows the word clouds obtained from CrimeBB posts grouped by PoC, exploitation, and weaponization classes and across  all posts. The presence of the word ``multisploit'' in the PoC category indicates that Trillium Security MultiSploit Tool is a popular tool used by hackers. 
The presence of the word  ``thanks'', in the exploitation cloud, refers to the behavior of users that confirm that a given exploitation worked.

Note that the term ``clean'' is also prevalent across posts about exploitation in the wild.  Indeed, it appears in the context of FUD exploits. Users typically indicate that they were able to run the exploit and that it was not detected by any antivirus. As an example, in  one post, we found 35 occurrences  of the word ``clean'' when referring to a   ``silent'' exploit to CVE-2011-3544.  
The exploit authors   share its MD5 hash and report a  
detection ratio of 0/35. To     evidence   the lack of detection, the authors present a  sample output:  \texttt{AVG Free: Clean; 
ArcaVir: Clean; 
Avast 5: Clean}. The list continues with  35 antivirus programs.

\section{Classifiers:  Distinguishing Potential Threat Against Eminent Threat} \label{sec:classifier}

\subsection{Feature extraction and Training of classifier}

To encode textual information, we consider three alternatives:  BoW, TF-IDF, and doc2vec. For BoW and TF-IDF, we consider the following four parameters: 
1)   the top 30,000 most frequently occurring words, 2) that appear at least 5 times, and 3) in at least 90\% of the posts  in the corpus are considered for analysis. In addition, we consider 4)    $n$-grams,  ranging from one word to three words.
%
%
%
%
%
%
%
For  doc2vec, we encode posts  into   5000-dimensional vectors.
 We use standard NLP  pre-processing techniques, e.g., filtering English language stop words and  punctuation   from the posts. Then, we proceed with the use of decision trees and random forests for classification. 


We started with an imbalanced dataset.  Recall from the last line of Table~\ref{tab:table_posts_citing_cves} that our dataset comprises 244 threads labeled as PoC, 397 as weaponization, and 102 as exploitation. To address this imbalance, we used the Random Over Sampling heuristic~\cite{over_sampler} to produce a balanced dataset. We split the dataset into 75\% for training+validation and 25\% for testing. We then conducted a grid search to find the optimal hyperparameters, such as tree depth, the number of features to consider at each tree split, minimum samples required to split an internal node, and maximum node degree. To maintain the proportion of categories during training, we employed 5-fold stratified cross-validation on the training+validation set.


\begin{table}[t]
    \caption{Decision Tree (DT) and Random Forest (RF) performance. \myhl{fill up} } 
    \centering  \label{tab:classeval}
   \scalebox{0.85}{    \begin{tabular}{l|l|l||c|c|c|c}
           & Text encoding &  Target classes & Accuracy & Precision & Recall & F1  \\
         \hline
         \hline
         DT & BoW & PoC, Weaponization, Exploitation & 0.71 & 0.71 & 0.72 & 0.70\\
         DT & TF-IDF & PoC, Weaponization, Exploitation & 0.73 & 0.73 & 0.74  & 0.72\\
         DT & doc2vec & PoC, Weaponization, Exploitation & 0.74 & 0.74 & 0.74 & 0.73\\ \hdashline
         
         DT & BoW & Exploitation vs Non-exploitation & 0.85 & 0.86 & 0.85 & 0.85 \\
         DT & TF-IDF & Exploitation vs Non-exploitation & 0.91 & 0.91 & 0.91 & 0.91 \\
         DT & doc2vec &Exploitation vs Non-exploitation& \textbf{0.92} & \textbf{0.93} & \textbf{0.92} & \textbf{0.92} \\ \hdashline
         
         DT & BoW & PoC vs Non-PoC & 0.75 & 0.75 & 0.75 & 0.75\\
         DT & TF-IDF & PoC vs Non-PoC &0.77 & 0.78 & 0.77 & 0.77  \\
         DT & doc2vec & PoC vs Non-PoC&   0.70 & 0.71 & 0.70 & 0.70 \\ \hdashline
         
         DT & BoW & Weaponization vs Non-weapon. &0.68 & 0.68 & 0.68 & 0.68\\
         DT & TF-IDF & Weaponization vs Non-weapon. & 0.63  &  0.64& 0.63 & 0.62\\
         DT & doc2vec &Weaponization vs Non-weapon.& 0.59 & 0.59 & 0.59 & 0.59\\ \hline \hline

         RF & BoW & PoC, Weaponization, Exploitation & 0.85 & 0.84 & 0.85 & 0.84\\
         RF & TF-IDF & PoC, Weaponization, Exploitation & 0.86 & 0.87 & 0.86  & 0.86\\
         RF & doc2vec & PoC, Weaponization, Exploitation &  0.86 & 0.90 & 0.86 & 0.86\\ \hdashline
         
         RF & BoW & Exploitation vs Non-exploitation & 0.98 & 0.98 & 0.98 & 0.98 \\
         RF & TF-IDF & Exploitation vs Non-exploitation & 0.98 & 0.98 & 0.98 & 0.98 \\
         RF & doc2vec &Exploitation vs Non-exploitation& \textbf{0.99} & \textbf{0.99} & \textbf{0.99} & \textbf{0.99} \\  \hdashline
         
         RF & BoW & PoC vs Non-PoC & 0.84 & 0.84 & 0.84 & 0.84\\
         RF & TF-IDF & PoC vs Non-PoC &0.87 & 0.87 & 0.87 & 0.87  \\
         RF & doc2vec & PoC vs Non-PoC&   0.88 & 0.90 & 0.88 & 0.87 \\
         \hdashline
         
         RF & BoW & Weaponization vs Non-weapon. &0.67 & 0.67 & 0.67 & 0.67\\
         RF & TF-IDF & Weaponization vs Non-weapon. & 0.69  &  0.70 & 0.69 & 0.69\\
         RF & doc2vec &Weaponization vs Non-weapon.& 0.58 & 0.58 & 0.58 & 0.58\\ 
         \hline
    \end{tabular} }
\end{table}

\subsection{Evaluation  and interpretation of classifiers}

\begin{figure}[t!]
    \centering
    \includegraphics[width=\columnwidth]{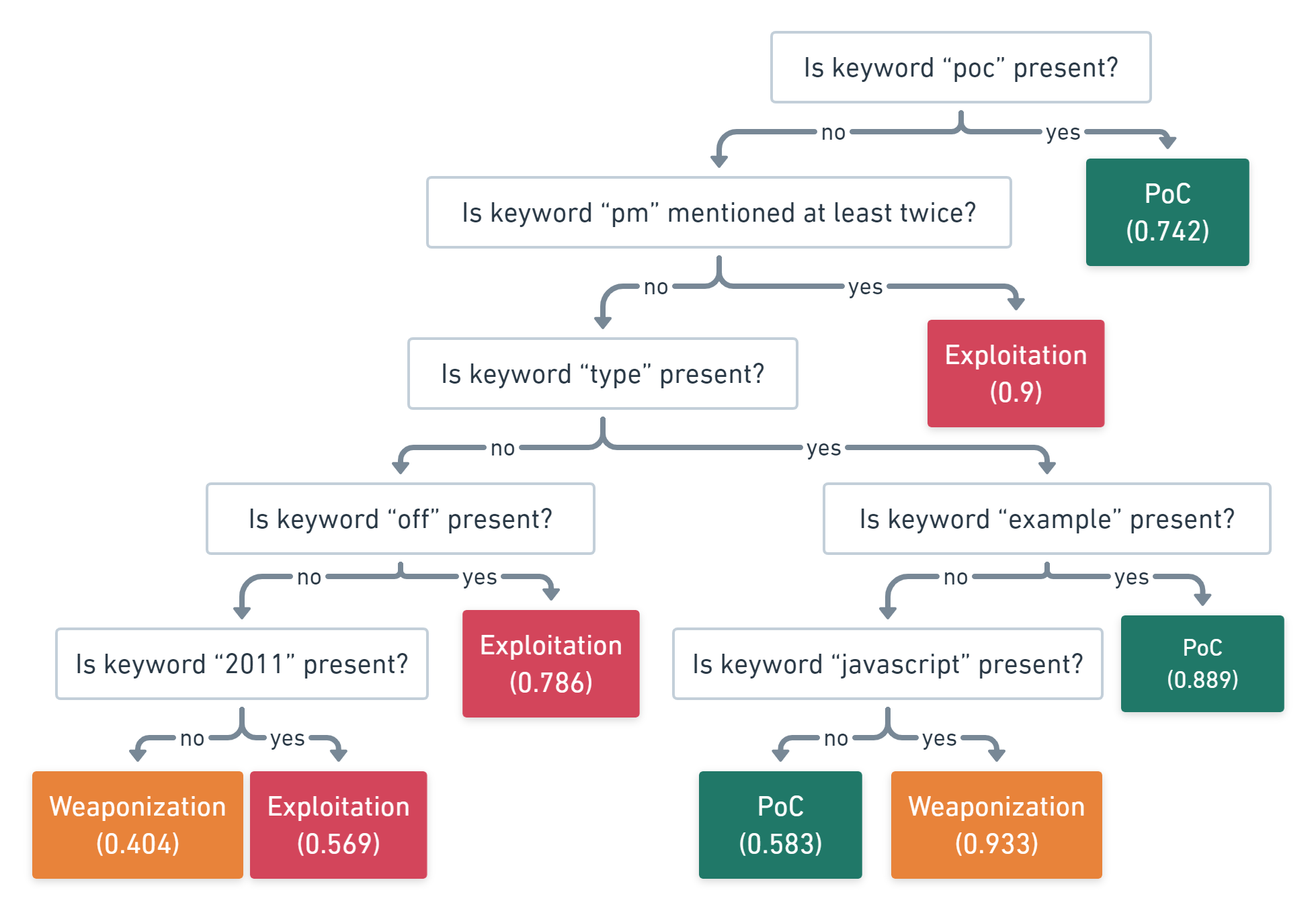}
    \caption{Decision tree   to classify PoC, weaponization, and  exploitation.}
    \label{fig:tree_3_classes_bow}  
\end{figure}

In this section, we report the performance of the considered classifiers. To that aim, we account for  four metrics: accuracy, precision, recall, and F1.  Table~\ref{tab:classeval} shows the obtained results, considering the best set of hyperparameters for each configuration,  as described above. In particular, our configurations vary as a function of the text encoding strategy and target classes.   
We observed increasing accuracy when switching from
the simpler and more interpretable encoding (BoW) to the
most complex but less interpretable one (doc2vec). Indeed,
doc2vec outperforms BoW and TF-IDF except for the cases
``PoC versus Non-PoC'' and ``Weaponization versus Non-
Weaponization''. Nonetheless,  BoW is instrumental to produce the interpretable tree presented in Figure~\ref{fig:tree_3_classes_bow}.

With respect to the target classes, we consider PoC vs Weaponization vs Exploitation and three additional     one-against-all classifiers, in which each  binary classifier  separates members of a class from members of other classes. The best results were obtained when filtering exploitation in the wild from the rest of the threads, which  is arguably the first step towards identifying relevant information at underground forums, as exploitation poses the most eminent risk. With respect to the classifier model, decision trees are simpler than random forests, producing less accurate predictions but being amenable to interpretation, as illustrated below.

Figure~\ref{fig:tree_3_classes_bow} illustrates the decision tree used to classify between PoC, weaponization, and exploitation (first line in Table~\ref{tab:classeval}).  Each internal node in the tree contains a rule that splits the dataset, and each leaf  indicates the most prominent class at that  split  and its frequency.  Despite the fact that not all splitting rules that appear in Figure~\ref{fig:tree_3_classes_bow} are interpretable, we can already extract interesting insights from it.  In the root of the tree, we find the rule with the highest splitting  power, according to the Gini index criterion.  Indeed, the root together with the leaf immediately below it indicate that if the thread contains the keyword ``poc'', with a 74.2\% chance,  it is actually a proof-of-concept. The  following rule indicates that posts wherein users are concerned about privacy, i.e., containing the keyword ``pm'', which stands for ``private message'' in the black forum jargon, correspond to exploitation in the wild.  Finally, we also observe that   JavaScript is a common language used to produce exploits, e.g., that inject code through unverified input fields.

\section{Conclusion} \label{sec:conclusion}

``Data is power as long as you know how to wield it.''  
%
 %
In this work, we  leveraged CrimeBB 
 and  machine learning methods to learn textual content and distinguish between:
(1) potential threat  (proof of concept), (2) eminent threat (weaponization), and (3) criminals chatting about a threat (exploitation in the wild). Among our  empirical findings obtained by relating   CVSS  and EPSS against CrimeBB,  
we found  that the most cited CVEs are typically related to higher risks and that it is feasible to automatically filter exploitation threads, with an accuracy above  99\%.
 
We believe that this work opens up   interesting avenues for future research, including the use of transformers such as ChatGPT to distill data from online forums, and the analysis of additional labels and elements, such as the maturity level of   discussions and its correlation against  EPSS  scores.  We also aim to expand the number of  posts considered in our study, accounting  for  vulnerabilities cited in  forums by their names, for  ``named vulnerabilities'', as opposed to CVE-ids, e.g., Bleichenbacher as opposed to CVE-2018-12404.

\section*{Acknowledgment}

 This project was   sponsored by CAPES, CNPq, and FAPERJ   (315110/2020-1, E-26/211.144/2019  and E-26/201.376/2021).

%
%
%
\bibliographystyle{splncs04}
\bibliography{mybibliography}
\end{document}